\documentclass[]{eptcs}
\providecommand{\event}{TARK 2021} 
\usepackage{breakurl}             
\usepackage{underscore}           
\usepackage[english]{babel}

\usepackage{amsthm, amsmath, amsfonts, amssymb}
\usepackage{comment}
\usepackage{ascmac}
\usepackage{mathtools}

\title{Are the Players in an Interactive Belief Model Meta-certain of the Model Itself? (Extended Abstract)}
\author{Satoshi Fukuda
\institute{Department of Decision Sciences and IGIER, Bocconi University\thanks{I would like to thank three anonymous referees of TARK 2021 for their helpful comments. This is an abbreviated technical summary of part of the full paper, which is available at the author's website (https://websfukuda.com/research/).}\\
Milan, Italy}
\email{satoshi.fukuda@unibocconi.it}
}
\def\titlerunning{Are the Players in an Interactive Belief Model Meta-certain of the Model Itself?}
\def\authorrunning{S. Fukuda}

\newtheorem{thm}{Theorem}
\newtheorem{prop}{Proposition} 

\theoremstyle{definition}
\newtheorem{defi}{Definition}

\newtheorem{rmk}{Remark}

\begin{document}
\maketitle

\begin{abstract}
In an interactive belief model, are the players \textquotedblleft commonly meta-certain'' of the model itself? This paper formalizes such implicit \textquotedblleft common meta-certainty'' assumption. To that end, the paper expands the objects of players' beliefs from events to functions defined on the underlying states. Then, the paper defines a player's belief-generating map: it associates, with each state, whether a player believes each event at that state. The paper formalizes what it means by: \textquotedblleft a player is (meta-)certain of her own belief-generating map'' or \textquotedblleft the players are (meta-)certain of the profile of belief-generating maps (i.e., the model).'' The paper shows: a player is (meta-)certain of her own belief-generating map if and only if her beliefs are introspective. The players are commonly (meta-)certain of the model if and only if, for any event which some player $i$ believes at some state, it is common belief at the state that player $i$ believes the event. This paper then asks whether the \textquotedblleft common meta-certainty'' assumption is needed for an epistemic characterization of game-theoretic solution concepts. The paper shows: if each player is logical and (meta-)certain of her own strategy and belief-generating map, then each player correctly believes her own rationality. Consequently, common belief in rationality alone leads to actions that survive iterated elimination of strictly dominated actions.
\end{abstract}

\section{Introduction}

In an economic or game-theoretic model in which the players make their interactive reasoning about their strategies or rationality, the analysts \textquotedblleft from outside of the model'' assume that the players \textquotedblleft commonly know'' the model itself. Since the pioneering work of \cite{Aumann76, Aumann87, Aumann99}, how to model such assumption and what consequences such assumption has have been puzzling economic and game theorists.\footnote{\label{fn:meta_knowledge}For this question, see also \cite{Bacharach85, Bacharach90}, \cite{BinmoreBrandenburger}, \cite{BDCK, BDHierarchies}, \cite{BDG}, \cite{BKImpossibility}, \cite{DekelGul}, \cite{FGHV}, \cite{GilboaMeta}, \cite{Pires}, \cite{RoyPacuit}, \cite{TanWerlangJET}, \cite{VassilakisZamir}, \cite{WerlangCKDict}, \cite{TanWerlangRBE}, and \cite{Wilson87}.}  

This paper has two objectives. The first aims at formalizing the \textquotedblleft common knowledge'' assumption of a model within the model itself. An interactive belief/knowledge model represents players' beliefs/knowledge about its ingredients, that is, events. The model itself does not tell whether the players (commonly) believe/know the model itself, although the analysts assume that the players (commonly) believe/know the model in a meta-sense. I refer to the knowledge/belief of the model as the \textquotedblleft meta-knowledge/meta-belief'' of the model.\footnote{\label{fn:metacertainty}Since different epistemic models may feature different notions of qualitative or probabilistic beliefs or knowledge, I use the word the \textquotedblleft (meta-)certainty'' of a model to refer generically to the meta-knowledge or meta-belief of the model.} 

The second objective is to examine the role that \textquotedblleft meta-knowledge'' of a model plays in game-theoretic analyses such as epistemic characterizations of solution concepts. For a given epistemic characterization of a game-theoretic solution concept such as iterated elimination of strictly dominated actions, do the outside analysts need to formally assume that the players \textquotedblleft meta-know'' an epistemic model of a game (that describes their interactive beliefs about their strategies and rationality)?  

The first main result, presented in Section \ref{sec:ckmodel}, characterizes the \textquotedblleft common meta-certainty'' assumption as follows. According to the formal test to be discussed, the players are commonly (meta-)certain of a model if and only if, for any event which some player $i$ believes at some state, it is common belief that player $i$ believes the event at that state. 

In Section \ref{sec:setup}, I start with introducing a (belief) model. The model consists of the following three ingredients. The first is a measurable space of states of the word $(\Omega, \mathcal{D})$. Each state $ \omega \in \Omega$ is a list of possible specifications of the world, and the collection $\mathcal{D}$ of events (i.e., subsets of $\Omega$) are the objects of the players' beliefs. The second is the players' monotone belief operators $(B_{i})_{i \in I}$. Player $i$'s belief operator $B_{i}$ associates, with each event $E$, the event that she believes $E$. Monotonicity means that if player $i$ believes $E$ at a state and if $E$ implies (i.e., is included in) $F$, then she believes $F$ at that state. The third is a common belief operator $C$, which associates, with each event $E$, the event that the players commonly believe $E$. Under certain assumptions on the players' beliefs, an event $E$ is common belief if and only if everybody believes $E$, everybody believes that everybody believes $E$, and so on \textit{ad infinitum}.

The framework nests the following two standard models of belief or knowledge (or combinations thereof). The first is a possibility correspondence model of qualitative belief or knowledge  (e.g., \cite{Aumann76, Aumann99}, \cite{DekelGul}, \cite{Geanakoplos}, and \cite{MorrisJET}). The possibility correspondence associates, with each state, the set of states that she considers possible. The player believes an event $E$ at a state when the possibility set at $\omega$  is included in the event $E$. The second is a type space (\cite{Harsanyi}), where each player's probabilistic beliefs are induced by her type mapping. The type mapping $\tau_{i}$ associates, with each state $\omega$, her probability measure $\tau_{i}(\omega)$ on the underlying states at that state. The type mapping $\tau_{i}$ of player $i$ induces her $p$-belief operator (\cite{MondererSamet}): it associates, with each event $E$, the event that player $i$ believes $E$ with probability at least $p$. 

With the framework in mind, I formalize the (meta-)certainty of a model in two steps. In the first step, Section \ref{sec:knowledgefunction} expands the objects of the players' beliefs from events to functions defined on the underlying states. Examples of such functions are random variables, strategies, and type mappings. Any such function $x$ has to be defined on the state space $\Omega$, but the co-domain $X$ can be any set such as the set $\mathbb{R}$ of real numbers (a random variable), a set $A_{i}$ of player $i$'s actions (her strategy), and the set $\Delta(\Omega)$ of probability measures on $(\Omega, \mathcal{D})$ (a type mapping). I call the function $x: \Omega \rightarrow X$ a signal if its co-domain $X$ has \textquotedblleft observational'' contents $\mathcal{X}$ (where \textquotedblleft observation'' is broadly construed as being an object of reasoning): it is a collection of subsets of $X$ such that each $F \in \mathcal{X}$ is deemed an event $x^{-1}(F)$. Formally, a \textit{signal (mapping)} is a function $x: (\Omega, \mathcal{D}) \rightarrow (X, \mathcal{X})$ such that each observation $F \in \mathcal{X}$ is considered to be an event $x^{-1}(F) \in \mathcal{D}$. Player $i$ is \textit{certain} of the value of the signal $x$ at a state $\omega$ if, for any observational content $F$ that holds at $\omega$ (i.e., $\omega \in x^{-1}(F)$), player $i$ believes the event $x^{-1}(F)$ at $\omega$ (i.e., $\omega \in B_{i}(x^{-1}(F))$). Player $i$ is \textit{certain} of $x$ if she is certain of the value of $x$ at every state. For example, let $x: \Omega \rightarrow X$ be the strategy of player $i$ and let every singleton action $a \in X$ be observable to her; then, player $i$ is certain of her own strategy if, wherever she takes an action $a=x(\omega)$ at a state $\omega$, she believes at $\omega$ that she takes action $a$. Having defined individual players' (meta-)certainty, the players are \textit{commonly certain} of the value of the signal $x$ at a state $\omega$ if, for any observational content $F$ that holds at $\omega$, the event $x^{-1}(F)$ is common belief at $\omega$ (i.e., $\omega \in C(x^{-1}(F))$). The players are \textit{commonly certain} of the signal $x$ if they are commonly certain of its value at every state. 

In the second step, Section \ref{sec:typerep} formulates a players' \textquotedblleft belief-generating map'' as a signal that associates, with each state, her beliefs at that state. By the second step, I can apply the formalization of certainty and common certainty in the first step to the ingredients of a given model (i.e., players' belief-generating maps). To that end, take player $i$'s belief operator $B_{i}$ from the model. I define a qualitative-type mapping $t_{B_{i}}$: it associates, with each state, whether player $i$ believes each event or not at that state (formally, a binary set function from the collection of events to the binary values $\{ 0,1 \}$ where $1$ indicates the belief of an event). The qualitative-type mapping is a binary \textquotedblleft type'' mapping analogous to a type mapping $\tau_{i}$ that represents player $i$'s probabilistic beliefs at each state in the context of probabilistic beliefs. Thus, the qualitative-type mapping $t_{B_{i}}$ associates, with each state $\omega$, her qualitative belief $t_{B_{i}}(\omega) \in \{ 0,1 \}$ (where $t_{B_{i}}(\omega)(E)=1$ if and only if $\omega \in B_{i}(E)$) on $(\Omega, \mathcal{D})$ at $\omega$. The qualitative-type mapping $t_{B_{i}}$ is player $i$'s belief-generating mapping. Since the belief operator $B_{i}$ and the qualitative-type mapping $t_{B_{i}}$ are equivalent means of representing player $i$'s beliefs,  a model means the profile of qualitative-type mappings. Thus, the formal test for whether the players are commonly certain of a given belief model is whether the players are certain of the profile of their qualitative-type mappings.  

Before asking when a player is certain of all the players' qualitative-type mappings (i.e., the model), Section \ref{sec:signal} characterizes when a player is certain of her own qualitative type-mapping in terms of her introspective properties of beliefs. Roughly, Proposition \ref{prop:CertaintyIntrospection} shows that each player is certain of her own qualitative-type mapping if and only if her belief is introspective. 

To the best of my knowledge, this is the first paper which systematically formalizes the statement that the players are (commonly) (meta-)certain of  any given belief model within the model itself. The main result on this question, nevertheless, is related to \cite{GilboaMeta}. He constructs a particular syntactic model in which the statement that the model is common knowledge is incorporated within itself. He formulates the sense in which the model is commonly known from Positive Introspection of common knowledge: if a statement is common knowledge then it is commonly known that the statement is common knowledge. In Theorem \ref{thm:common_certain_qualitative}, in contrast, the players are commonly certain of a given model if and only if, at each state and for any event which some player believes at that state, it is common belief that the player believes the event at that state. Thus, in this paper, the key criteria is the positive introspective property of common belief with respect to each player's beliefs. Whenever some individual player believes some event, it is common belief that she believes it. \cite{Bacharach85, Bacharach90}, in the context of partitional possibility correspondence models, formalizes the event that a player has an information partition by regarding it as a function.   

Having characterized the common-certainty of the model, Section \ref{sec:solutioncertainty} examines the role that the \textquotedblleft common meta-certainty'' assumption plays in the epistemic characterization of iterated elimination of strictly dominated actions (IESDA) in a strategic game. Informally, it states: if the players are \textquotedblleft logical,'' if they are commonly meta-certain of a game, and if they commonly believe their rationality, then the resulting actions survive any process (i.e., any order) of IESDA. Formally, it states: if the players commonly believe their rationality and if their common belief in their rationality is correct, then the resulting actions survive any process of IESDA.\footnote{The formal statement is taken from \cite[Theorem 3]{FukudaCB}, which holds irrespective of the nature of beliefs. For seminal papers on implications of common belief in rationality, see, for example, \cite{BrandenburgerDekelEcta87}, \cite{StalnakerTD94}, and \cite{TanWerlangJET}.} Theorem \ref{thm:informativeness_rationality} connects these two statements. If the players' beliefs are monotone (they believe any logical implication of their beliefs), consistent (i.e., they do not simultaneously believe an event and its negation), and finitely conjunctive (if they believe $E$ and $F$ then they believe its conjunction $E \cap F$), and if each player is certain of her own strategy and the part of her own belief-generating process in the model (each player is not necessarily certain of how the opponents' beliefs are generated in the model), then each player correctly believes her own rationality, and hence they have correct common belief in their rationality. Thus, if the players are \textquotedblleft logical'' and each of them is meta-certain of the part of the model that governs her own beliefs, then common belief in rationality leads to actions that survive any process of IESDA. 

On the one hand, Theorem \ref{thm:common_certain_qualitative} states that the players may not always be commonly certain of a belief model in the sense that each player is commonly certain of how the other players' beliefs are represented within the model. On the other hand, Theorem \ref{thm:informativeness_rationality} asserts that common belief in rationality leads to IESDA even if the players may not be commonly certain of the belief model. 

The paper is organized as follows. Section \ref{sec:setup} defines a belief model. Section \ref{sec:structureknowledge} characterizes the sense in which each player is certain of how her belief is generated in a model. Section \ref{sec:ckmodel} examines the sense in which the players are commonly certain of a model itself (i.e., how the players' beliefs are generated in the model). Section \ref{sec:solutioncertainty} studies how the assumption that the players are commonly certain of a model itself can make game-theoretic analyses coherent. 

\section{Framework} \label{sec:setup}

Throughout the paper, let $I$ denote a non-empty finite set of \textit{players}. The framework represents players' interactive beliefs by belief operators on a state space. Section \ref{sec:beliefspace} defines a belief model. Section \ref{sec:propertiesbeliefs} defines properties of beliefs. 

\subsection{A Belief Model}\label{sec:beliefspace}

A \textit{belief model} is a tuple $\overrightarrow{\Omega} := \langle (\Omega, \mathcal{D}), (B_{i})_{i \in I}, C \rangle$, where: (i) $(\Omega, \mathcal{D})$ is a non-empty measurable space of states of the world (call $\Omega$ the \textit{state space}); (ii) $B_{i}: \mathcal{D} \rightarrow \mathcal{D}$ is player $i$'s (monotone) \textit{belief operator}; and (iii) $C: \mathcal{D} \rightarrow \mathcal{D}$ is a (monotone) \textit{common belief operator} to be defined below. 

While $\Omega$ constitutes a non-empty set of \textit{states} of the world, each element $E$ of $\mathcal{D}$ is an \textit{event} about which the players reason. For each event $E$, the set $B_{i}(E)$ denotes the event that (i.e., the set of states at which) a player $i$ believes $E$. Thus, the player $i \in I$ believes an event $E \in \mathcal{D}$ at a state $\omega \in \Omega$ if $\omega \in B_{i}(E)$. I assume that each player's belief operator satisfies \textit{Monotonicity}: $E \subseteq F$ implies $B_{i}(E) \subseteq B_{i}(F)$. It means that if player $i$ believes some event then she believes any of its logical consequences.\footnote{The full paper dispenses with the monotonicity assumption, which enables one to define the meta-certainty of any belief model.} 

Since the players' beliefs are monotone, I introduce the common belief operator $C: \mathcal{D} \rightarrow \mathcal{D}$ following \cite{MondererSamet}. Call an event $E$ \textit{publicly evident} if $E \subseteq B_{I}(E) :=\bigcap_{i \in I} B_{i}(E)$. That is, everybody believes $E$ whenever $E$ is true. Denote by $\mathcal{J}_{B_{I}}$ the collection of publicly-evident events. An event $E$ is \textit{common belief} at a state $\omega$ if there is a publicly-evident event that is true at $\omega$ and that implies the mutual belief in $E$: that is, $\omega \in F \subseteq B_{I}(E)$ for some $F \in \mathcal{J}_{B_{I}}$. Now, $C$ is assumed to satisfy that the set of states at which $E$ is common belief is an event for each $E \in \mathcal{D}$: $C(E):=\{ \omega \in \Omega \mid \text{there is } F \in \mathcal{J}_{B_{I}} \text{ with } \omega \in F \subseteq B_{I}(E) \} \in \mathcal{D}$.

Since players' beliefs are monotone and since $\mathcal{D}$ is closed under countable intersection, if $E$ is common belief, then  everybody believes $E$, everybody believes that everybody believes $E$, and so forth \textit{ad infinitum}: $C(E) \subseteq \bigcap_{n \in \mathbb{N}} B^{n}_{I}(E)$. The converse (set inclusion) holds, for example, when the mutual belief operator $B_{I}$ (or every $B_{i}$) satisfies \textit{Countable Conjunction}: $\bigcap_{n \in \mathbb{N}} B_{I}(E_{n}) \subseteq B_{I}(\bigcap_{n \in \mathbb{N}} E_{n} )$, meaning that everybody believes the countable conjunction of events whenever everybody believes each of them.  

I represent the players' beliefs on a measurable space $(\Omega, \mathcal{D})$ so that I can analyze players' qualitative and probabilistic beliefs  under the same framework. The full paper introduces the players'  probabilistic beliefs on a measurable space $(\Omega, \mathcal{D})$ by $p$-belief operators \cite{MondererSamet}. For each $p \in [0,1]$, player $i$'s $p$-belief operator $B^{p}_{i}$ associates, with each event $E$, the event that player $i$ believes $E$ with probability at least $p$. The full paper also introduces the common $p$-belief operator $C^{p}$. 

\subsection{Properties of Beliefs}\label{sec:propertiesbeliefs}

Next, I introduce additional eight properties of beliefs. Various possibility correspondence models of qualitative beliefs and knowledge  are represented as belief models that satisfy certain properties specified below. Fix a player $i$. I first introduce the following five logical properties of beliefs.
\begin{enumerate}
\item \textit{Necessitation}: $B_{i}(\Omega) = \Omega$. 
\item \textit{Countable Conjunction}: $\bigcap_{n \in \mathbb{N}} B_{i}(E_{n}) \subseteq B_{i} ( \bigcap_{n \in \mathbb{N}}E_{n} )$ (for any events $(E_{n})_{n \in \mathbb{N}}$).
\item \textit{Finite Conjunction}: $B_{i}(E) \cap B_{i}(F) \subseteq B_{i}(E \cap F)$.
\item The \textit{Kripke property}: $B_{i}(E) = \{ \omega \in \Omega \mid b_{B_{i}}(\omega) \subseteq E \}$, where $b_{B_{i}}(\omega) :=\bigcap \{ E \in \mathcal{D} \mid \omega \in B_{i}(E) \}$ is the set of states player $i$ considers \textit{possible} at $\omega$. 
\item \textit{Consistency}: $B_{i}(E) \subseteq (\neg B_{i})(E^{c})$.
\end{enumerate}
First, Necessitation means that the player believes a tautology such as $E \cup E^{c}$. Second, as discussed, Countable Conjunction means that if the player believes each of a countable collection of events, then she believes its conjunction. Third, Finite Conjunction is weaker than Countable Conjunction: if the player believes $E$ and $F$ then she believes its conjunction $E \cap F$. Fourth, to discuss the Kripke property, the player considers $\omega'$ possible at $\omega$ if, for any event $E$ which she believes at $\omega$, $E$ is true at $\omega'$. The Kripke property provides the condition under which $i$'s belief is induced by her \textit{possibility correspondence} $b_{B_{i}}: \Omega \rightarrow \mathcal{P}(\Omega)$: she believes $E$ at $\omega$ if and only if (hereafter, iff) her possibility set $b_{B_{i}}(\omega)$ at $\omega$ implies $E$. The Kripke property implies the previous three properties as well as Monotonicity. Fifth, Consistency means that the player cannot simultaneously believe an event $E$ and its negation $E^{c}$.

Next, I move on to truth and introspective properties. 
\begin{enumerate}\setcounter{enumi}{5}
\item \textit{Truth Axiom}: $B_{i}(E) \subseteq E$ (for all $E \in \mathcal{D}$).
\item \textit{Positive Introspection}: $B_{i}(\cdot) \subseteq B_{i}B_{i}(\cdot)$ (i.e., $B_{i}(E) \subseteq B_{i}B_{i}(E)$ for all $E \in \mathcal{D}$).
\item \textit{Negative Introspection}: $(\neg B_{i})(\cdot) \subseteq B_{i}(\neg B_{i})(\cdot)$.
\end{enumerate}
Sixth, Truth Axiom turns belief into knowledge in that knowledge has to be true while belief can be false. Truth Axiom implies Consistency. While knowledge satisfies Truth Axiom, qualitative and probabilistic beliefs are often assumed to satisfy Consistency. Seventh, Positive Introspection states that if the player believes some event then she believes that she believes it. Eighth, Negative Introspection states that if the player does not believe some event then she believes that she does not believe it. Truth Axiom and Negative Introspection yield Positive Introspection (e.g., \cite{Aumann99}).

Three remarks are in order. First, the introspective properties will play important roles in whether a player is (meta-)certain of a belief model. Intuitively, Positive Introspection provides the sense in which the player believes her own belief (at least at face value) while Negative Introspection yields the sense in which the player believes the lack of her own belief. To see these points, an event $E$ is \textit{self-evident} to $i$ if $E \subseteq B_{i}(E)$. That is, $i$ believes $E$ whenever $E$ is true. Positive Introspection means that $i$'s belief in $E$ is self-evident to $i$, and Negative Introspection means that $i$'s lack of belief in $E$ is self-evident to $i$. Denote by $\mathcal{J}_{B_{i}}$ the collection of self-evident events to $i$.

Second, the last four properties are restated in terms of $b_{B_{i}}$ under the Kripke property: $B_{i}$ satisfies Consistency iff $b_{B_{i}}$ is serial (i.e., $b_{B_{i}}(\cdot) \neq \emptyset$); $B_{i}$ satisfies Truth Axiom iff $b_{B_{i}}$ is reflexive (i.e., $\omega \in b_{B_{i}}(\omega)$ for all $\omega \in \Omega$); $B_{i}$ satisfies Positive Introspection iff $b_{B_{i}}$ is transitive (i.e., $\omega' \in b_{B_{i}}(\omega)$ implies $b_{B_{i}}(\omega') \subseteq b_{B_{i}}(\omega)$); and $B_{i}$ satisfies Negative Introspection iff $b_{B_{i}}$ is Euclidean (i.e., $\omega' \in b_{B_{i}}(\omega)$ implies $b_{B_{i}}(\omega) \subseteq b_{B_{i}}(\omega')$). 

Third, various models of probabilistic and qualitative beliefs and knowledge take different sets of axioms. The framework accommodates possibility correspondence models of qualitative beliefs and knowledge when $B_{i}$ satisfies the Kripke property. A partitional model of knowledge corresponds to the case when $B_{i}$ satisfies Truth Axiom, Positive Introspection, and Negative Introspection.\footnote{In fact, Truth Axiom, Negative Introspection, and the Kripke property yield all the other properties defined in this section.} A reflexive and transitive (non-partitional) possibility correspondence model is characterized by Truth Axiom and Positive Introspection.\footnote{\label{fn:non_partitional}See, for example, \cite{Bacharach85}, \cite{BinmoreBrandenburger}, \cite{DekelGul}, \cite{Geanakoplos}, \cite{Pires}, \cite{SametIgnorance}, and \cite{Shin}.} When it comes to fully-introspective qualitative beliefs, $b_{B_{i}}$ is serial, transitive, and Euclidean iff $B_{i}$ satisfies Consistency, Positive Introspection, and Negative Introspection. 



\section{When Is a Player Certain of Her Belief-Generating Mapping?} \label{sec:structureknowledge}

Section \ref{sec:knowledgefunction} extends an object of beliefs in a model from an event to a function (\textquotedblleft signal'') defined on the state space. That is, the subsection formulates the statement that a player is certain of a function defined on the state space.  Section \ref{sec:typerep} represents a player's \textquotedblleft belief-generating mapping'' as a signal which associates, with each state, whether she believes each event or not.  Section \ref{sec:signal} asks the sense in which she is certain of her own belief-generating mapping in terms of the introspective properties. 

\subsection{Functions as Objects of Players' Beliefs} \label{sec:knowledgefunction}

I start with defining a notion of a signal mapping. A signal mapping is any function $x$ defined on the state space $\Omega$ with \textquotedblleft observational'' contents. A signal is interpreted as a mapping from the underlying state space $\Omega$ into a space of \textquotedblleft observation'' $X$ endowed with \textquotedblleft observational'' contents $\mathcal{X} \subseteq \mathcal{P}(X)$. By observation, it means that each $F \in \mathcal{X}$ is deemed an object of reasoning. That is, we call a mapping $x: \Omega \rightarrow X$ a signal mapping if each \textquotedblleft observational'' content $F \in \mathcal{X}$ can be regarded as an event $x^{-1}(F) \in \mathcal{D}$ through inverting the mapping. 

Formally, for a non-empty set $X$ and a non-empty subset $\mathcal{X}$ of $\mathcal{P}(X)$, call $x: (\Omega, \mathcal{D}) \rightarrow (X, \mathcal{X})$ a \textit{signal} (mapping) if $x^{-1}(\mathcal{X}) \subseteq \mathcal{D}$. Mathematically, $x: (\Omega, \mathcal{D}) \rightarrow (X, \mathcal{X})$ is a signal if $x: (\Omega, \mathcal{D}) \rightarrow (X, \sigma (\mathcal{X}))$ is measurable. Examples include strategies, random variables, and so on. 

The main purpose of this subsection is to define the statement that a player is certain of a signal. A player $i$ is \textit{certain of the value of} a signal $x: (\Omega, \mathcal{D}) \rightarrow (X, \mathcal{X})$ at $\omega$, if she believes any observational content $F$ (i.e., believes $x^{-1}(F)$) at $\omega$ whenever it is true: $x(\omega) \in F$. She is \textit{certain of} the signal $x: (\Omega, \mathcal{D}) \rightarrow (X, \mathcal{X})$ if she is certain of its value at every $\omega$. Likewise, the players are \textit{commonly certain of the value of} the signal $x: (\Omega, \mathcal{D}) \rightarrow (X, \mathcal{X})$ at $\omega$, if the players commonly believe any observational content $F$ at $\omega$ whenever it is true. The players are \textit{commonly certain of} the signal $x: (\Omega, \mathcal{D}) \rightarrow (X, \mathcal{X})$ if they are certain of its value at every $\omega$. 
Formally: 

\begin{defi}\label{df:signal} 
Let $\overrightarrow{\Omega}$ be a belief model, and let $x: (\Omega, \mathcal{D}) \rightarrow (X, \mathcal{X})$ be a signal mapping. 
\begin{enumerate}
\item \begin{enumerate}
\item Player $i$ is \textit{certain of the value of the signal} $x: (\Omega, \mathcal{D}) \rightarrow (X, \mathcal{X})$ at $\omega$ if $\omega \in B_{i}(x^{-1}(F))$ for any $F \in \mathcal{X}$ with $x(\omega) \in F$.
\item Player $i$ is \textit{certain of the signal} $x: (\Omega, \mathcal{D}) \rightarrow (X, \mathcal{X})$ if she is certain of the value of the signal $x$ at any state. 
\end{enumerate}
\item \begin{enumerate}
\item The players are \textit{commonly certain of the value of the signal} $x: (\Omega, \mathcal{D}) \rightarrow (X, \mathcal{X})$ at $\omega$ if $\omega \in C(x^{-1}(F))$ for any $F \in \mathcal{X}$ with $x(\omega) \in F$. 
\item The players \textit{are commonly certain of} the signal $x: (\Omega, \mathcal{D}) \rightarrow (X, \mathcal{X})$ if they are commonly certain of the value of the signal $x$ is at every state.
\end{enumerate}
\end{enumerate}
\end{defi}

For example, suppose that $x: \Omega \rightarrow X$ is a decision function of a player which associates, with each state, the action taken at that state. Suppose the set of actions $X$ is endowed with the collection of singleton actions $\mathcal{X} = \{ \{ a \} \mid a \in X \}$. Each action $a$ corresponds to an observational content to the player, and $x$ is a signal mapping if the set of states at which the player takes action $a$ is an event: $x^{-1}(\{ a \}) = \{ \omega \in \Omega \mid x(\omega)=a \} \in \mathcal{D}$ for each $a \in X$. 

More specifically, let $\Omega = \{ \omega_{1}, \omega_{2}, \omega_{3} \}$ and $X = \{ a,b \}$. For each $i \in I = \{ 1,2\}$, let $B_{i}$ be given by (i) $B_{i}(E) = E \setminus \{ \omega_{3} \}$ for each $E \neq \Omega$; and (ii) $B_{i}(\Omega) = \Omega$. Suppose player $1$'s decision function $x: (\Omega, \mathcal{P}(\Omega)) \rightarrow ( X, \{ \{ a\}, \{ b \} \} )$ is given by $(x(\omega))_{\omega \in \Omega} = (a,a,a)$. Since $B_{1}(\Omega) = \Omega$ and $B_{1}(\emptyset) = \emptyset$, whenever player $1$ takes a certain action, she believes that she takes that action. Thus, player $1$ is certain of $x$. If, instead, her decision function $x: (\Omega, \mathcal{P}(\Omega)) \rightarrow ( X, \{ \{ a\}, \{ b \} \} )$ is given by $(x(\omega))_{\omega \in \Omega} = (a,b,a)$, then at $\omega_{3}$ at which she takes action $a$, she does not believe that she takes action $a$, because $B_{1}(\{ \omega_{1}, \omega_{3} \}) = \{ \omega_{1} \}$. Thus, player $1$ is not certain of the value of $x$ at $x_{3}$. Since $C = B_{1}$, the same arguments hold for the common certainty of $x$.

For ease of terminology, player $i$ is \textit{certain of (the value of) the signal} $x: \Omega \rightarrow X$ (at $\omega$) \textit{with respect to} $\mathcal{X}$ if she is certain of (the value of) the signal $x: (\Omega, \mathcal{D}) \rightarrow (X, \mathcal{X})$ (at $\omega$). Likewise, the players are \textit{commonly certain of (the value of) the signal} $x: \Omega \rightarrow X$ (at $\omega$) \textit{with respect to} $\mathcal{X}$ if they are commonly certain of (the value of) the signal $x: (\Omega, \mathcal{D}) \rightarrow (X, \mathcal{X})$ (at $\omega$).

Four remarks on Definition \ref{df:signal} are in order. First, I restate the fact that a player is certain of a signal in terms of self-evidence. Namely, player $i$ is certain of a signal $x: (\Omega, \mathcal{D}) \rightarrow (X, \mathcal{X})$ iff any observational content $F \in \mathcal{X}$ (i.e., any event $x^{-1}(F) \in \mathcal{D}$) is self-evident to $i$. Likewise, the players are commonly certain of the signal $x: (\Omega, \mathcal{D}) \rightarrow (X, \mathcal{X})$ iff any observational content $F \in \mathcal{X}$ is publicly-evident. Consequently, the players are commonly certain of a signal iff every player is certain of it.\footnote{In contrast, it is not necessarily the case that each player is certain of a signal at a state $\omega$ iff the players are commonly certain of the signal at $\omega$.} 

Second, when $x: (\Omega, \mathcal{D}) \rightarrow (X, \mathcal{X})$ is a player's strategy, Definition \ref{df:signal} formalizes the statement that the player is certain of the strategy (e.g., \cite{BDG} and \cite{Geanakoplos}). To see this, assume that $\mathcal{X}$ contains a singleton $\{ x(\omega) \}$ to reason about the action taken at $\omega$. That is, the set of states $[x(\omega)]:= x^{-1}(\{ x(\omega) \}) = \{ \omega' \in \Omega \mid x(\omega') = x(\omega) \}$ at which player $i$ takes the same action as she does at $\omega$ is an event. Then, player $i$ is certain of her strategy iff $[x(\omega)]$ is self-evident at every $\omega \in \Omega$. In words, player $i$ is certain of her strategy $x$ iff, whenever she takes action $a=x(\omega)$ at $\omega$, she believes at $\omega$ that she takes action $a = x(\omega)$.

Definition \ref{df:signal} also subsumes the formulation of the certainty of the strategy by \cite{Aumann87} in the (countable) partitional state space model of knowledge. Let $(b_{B_{i}}(\omega))_{\omega \in \Omega}$ be a countable partition on $\Omega$. In \cite{Aumann87}, the player \textquotedblleft knows'' her own strategy $x$ iff the strategy $x$ is measurable with respect to the partition (which turns out to be equivalent to 
$b_{B_{i}}(\cdot) \subseteq [x(\cdot)]$). Since the partition is countable, the $\sigma$-algebra generated by the partition is equal to the self-evident collection: $\mathcal{J}_{B_{i}} = \sigma (\{ b_{B_{i}}(\omega) \in \mathcal{D} \mid \omega \in \Omega \})$. Hence, player $i$ is certain of her strategy $x: (\Omega, \mathcal{D}) \rightarrow (X, \mathcal{X})$ iff $x: (\Omega, \mathcal{J}_{B_{i}}) \rightarrow (X, \sigma(\mathcal{X}))$ is measurable.

Third, player $i$ satisfies Necessitation iff she is certain of any constant signal. Likewise, the common belief operator $C$ satisfies Necessitation (equivalently, every $B_{i}$ satisfies Necessitation) iff the players are commonly certain of any constant signal.

Necessitation allows the players to be certain of any constant \textquotedblleft random'' variable that does not depend on the realization of a state. For example, consider whether player $i$ is certain that an event $B_{j}(E)$ is equal to an event $F$ in a belief model. The outside analysts determine whether player $i$ believes that player $j$ believes an event $E$ at a state $\omega$ by examining whether $\omega \in B_{i}B_{j}(E)$ since player $j$'s belief $B_{j}(E)$ itself is an event. The (implicit) assumption in any (semantic) belief model is that $E=F$ implies $B_{i}(E) = B_{i}(F)$. Thus, if two events are extensionally the same, then each player's belief in the two events are the same.\footnote{\label{fn:syntaxcanonical} Although such identification of events are implicitly assumed for any (semantic) belief model, one can construct a canonical (\textquotedblleft universal'') semantic model from a syntactic language which maximally distinguishes the denotations of events. In the canonical model, such identification of events can be minimized in a way such that two events are equated only when they are explicitly assumed to be equivalent by the outside analysts (see \cite{FukudaUQB} for a formal assertion).} To assess player $i$'s belief about player $j$'s belief about $E$, how can the outside analysts justify the fact that player $i$ is able to equate $B_{j}(E)$ with another event (say, $F$)? Since either $B_{j}(E) = F$ or $B_{j}(E) \neq F$,  player $i$ is \textit{certain that $B_{j}(E)$ is an event $F$} if player $i$ is certain of the indicator function $\mathbb{I}_{B_{j} \leftrightarrow F}$, where $(B_{j}(E) \leftrightarrow F):=((\neg B_{j})(E) \cup F) \cap ((\neg F) \cup B_{j}(E))$. If player $i$'s belief operator $B_{i}$ satisfies Necessitation and if $B_{j}(E)=F$, then player $i$ is certain of the constant indicator function $\mathbb{I}_{B_{j} \leftrightarrow F}$. Thus, under Necessitation, player $i$ is certain that $B_{j}(E)=F$ if it is indeed the case. This argument justifies that, under Necessitation, the outside analysts can say that the players are certain of equating two extensionally equivalent events (say, $B_{j}(E)$ and $F$) if they are indeed extensionally equivalent. 

Fourth, it can be formally shown that player $i$ is certain of a profile of signals (e.g., a strategy profile) iff she is certain of each of them. Thus, the players are commonly certain of a profile of signals iff every player is certain of every signal.

\subsection{A Qualitative-Type Mapping that Represents a Player's Beliefs}\label{sec:typerep}

In order to formulate a test under which the outside analysts can examine whether the players are commonly certain of a belief model, I define the \textquotedblleft belief-generating map,'' which I call the qualitative-type mapping (\cite{{FukudaKnowledgeChap1}}), of a player. Given the belief operator of the player, the qualitative-type mapping associates, with each state, a binary value indicating whether the player believes each event in an analogous manner to the type mapping in the type-space literature.

To that end, recall that a (probabilistic-)type mapping associates, with each state $\omega$, the player's probabilistic beliefs $\tau_{i}(\omega) \in \Delta(\Omega)$ at that state. With this in mind, let $M(\Omega)$ be the set of binary set functions $\mu: \mathcal{D} \rightarrow \{ 0,1 \}$ (i.e., $M(\Omega) \subseteq \{ 0,1 \}^{\mathcal{D}}$) that satisfy a given set of logical properties of beliefs defined in Section \ref{sec:propertiesbeliefs} (these properties will be shortly expressed in terms of $\mu$). Call each $\mu \in M(\Omega)$ a \textit{qualitative-type}. Interpret $\mu (E) =1$ as the belief in an event $E \in \mathcal{D}$. Once $M(\Omega) \subseteq \{ 0,1 \}^{\mathcal{D}}$ is defined as the set of qualitative-types that satisfy the given set of logical properties of beliefs, I represent player $i$'s beliefs by a \textit{qualitative-type mapping} $t_{i}: \Omega \rightarrow M(\Omega)$ satisfying a certain measurability condition specified below. It is a measurable mapping which associates, with each state $\omega \in \Omega$, player $i$'s qualitative-type $t_{i}(\omega) \in M(\Omega)$ at $\omega$. Thus, player $i$ believes an event $E$ at $\omega$ if $t_{i}(\omega) (E)=1$. 

Now, I define the logical properties of $\mu$ in an analogous way to the corresponding logical properties of belief operators. Fix $\mu \in \{ 0,1 \}^{\mathcal{D}}$. 
\begin{enumerate}\setcounter{enumi}{-1}
\item \textit{Monotonicity}: $E \subseteq F$ implies $\mu (E) \leq \mu (F)$.
\item \textit{Necessitation}: $\mu(\Omega)=1$.
\item \textit{Countable Conjunction}: $\min_{n \in \mathbb{N}} \mu (E_{n}) \leq \mu ( \bigcap_{n \in \mathbb{N}} E_{n} )$.
\item \textit{Finite Conjunction}: $\min ( \mu (E), \mu(F)) \leq \mu ( E \cap F )$.
\item The \textit{Kripke property}: $\mu(E)=1$ iff $\bigcap \{ F \in \mathcal{D} \mid \mu (F) =1 \} \subseteq E$.
\item \textit{Consistency}: $\mu (E) \leq 1- \mu (E^{c})$. 
\end{enumerate}
The interpretations of the above properties are similar to those in Section \ref{sec:propertiesbeliefs}. 
Whether all of these properties are assumed or not depend on the model that the outside analysts study. For example, if the outside analysts examine a partitional possibility correspondence model, then $M(\Omega)$ is the set of qualitative-types that satisfy all the logical properties. 

I formally define the measurability condition of a qualitative-type mapping. A \textit{qualitative-type mapping} is a measurable mapping $t_{i}: (\Omega, \mathcal{D}) \rightarrow (M(\Omega), \mathcal{D}_{M})$ which satisfies given (logical and) introspective properties of beliefs, where $\mathcal{D}_{M}$ is the $\sigma$-algebra generated by the sets of the form $\beta_{E} := \{ \mu \in M (\Omega) \mid \mu (E)=1  \}$ for all $E \in \mathcal{D}$. The set $\beta_{E}$ is the set of types under which $E$ is believed. Thus, $\beta_{E}$ is an informational content indicating that event $E$ is believed. Note that $t_{i}: \Omega \rightarrow M(\Omega)$, by construction, satisfies given logical properties because any element in $M(\Omega)$ satisfies them. For example, if every $\mu \in M(\Omega)$ satisfies the Kripke property, then every $t_{i}(\omega)$ satisfies it. Denote $b_{t_{i}}(\omega):=\bigcap \{ E \in \mathcal{D} \mid t_{i}(\omega)(E)=1 \}$ for each $\omega \in \Omega$. 

The measurablity condition of $t_{i}$ requires each $t_{i}^{-1} ( \beta_{E} ) = \{ \omega \in \Omega \mid t_{i}(\omega)(E)=1 \}$ to be the event that player $i$ believes $E$. Next, I define Truth Axiom and the introspective properties of $t_{i}$.
\begin{enumerate}\setcounter{enumi}{5}
\item \textit{Truth Axiom}: $t_{i}(\omega)(E)=1$ implies $\omega \in E$.
\item \textit{Positive Introspection}: $t_{i}(\omega)(E)=1$ implies $t_{i}(\omega) (t_{i}^{-1} ( \beta_{E} )) =1$.
\item \textit{Negative Introspection}: $t_{i}(\omega)(E)=0$ implies $t_{i}(\omega) ( \neg t_{i}^{-1} ( \beta_{E} ) ) =1$.
\end{enumerate}


\subsection{Certainty of Own Type Mapping}\label{sec:signal}

I apply the certainty of a signal to a qualitative-type mapping. Proposition \ref{prop:CertaintyIntrospection} below roughly states that a player is certain of her own qualitative-type mapping iff her beliefs are introspective. 

\begin{prop}\label{prop:CertaintyIntrospection} 
Let $\overrightarrow{\Omega}$ be a belief model, and let $t_{B_{i}}: \Omega \rightarrow M(\Omega)$ be player $i$'s qualitative-type mapping.
\begin{enumerate}
\item \label{itm:prop_t_i_introspection_1} \begin{enumerate}
\item \label{itm:prop_t_i_introspection_1a} Player $i$ is certain of $t_{B_{i}}$ with respect to $\{ \beta_{E} \mid E \in \mathcal{D} \}$ iff  $B_{i}$ satisfies Positive Introspection. 
\item \label{itm:prop_t_i_introspection_1b} Player $i$ is certain of $t_{B_{i}}$ with respect to $\{ \neg \beta_{E} \mid E \in \mathcal{D} \}$ iff  $B_{i}$ satisfies Negative Introspection. 
\item \label{itm:prop_t_i_introspection_1c} If player $i$ is certain of $t_{B_{i}} : (\Omega, \mathcal{D}) \rightarrow (M(\Omega), \mathcal{D}_{M})$, then $B_{i}$ satisfies Positive Introspection and Negative Introspection.
\end{enumerate}
\item \label{itm:prop_t_i_introspection_2} 
\begin{enumerate}
\item \label{itm:prop_t_i_introspection_2a} Let $B_{i}$ satisfy Truth Axiom. Player $i$ is certain of $t_{B_{i}} : (\Omega, \mathcal{D}) \rightarrow (M(\Omega), \mathcal{D}_{M})$ iff $B_{i}$ satisfies (Positive Introspection and) Negative Introspection. 
\item \label{itm:prop_t_i_introspection_2b} Let $B_{i}$ satisfy Consistency and Countable Conjunction. Player $i$ is certain of $t_{B_{i}} : (\Omega, \mathcal{D}) \rightarrow (M(\Omega), \mathcal{D}_{M})$ iff $B_{i}$ satisfies Positive Introspection and Negative Introspection. 
\end{enumerate}
\end{enumerate}
\end{prop}

Part (\ref{itm:prop_t_i_introspection_1}) characterizes the certainty of the qualitative-type mapping $t_{B_{i}}$ with respect to the possession or lack of beliefs. In contrast, Parts (\ref{itm:prop_t_i_introspection_2a}) and (\ref{itm:prop_t_i_introspection_2b}), respectively,  examine the sense in which player $i$ is certain of her qualitative-type mapping $t_{B_{i}} : (\Omega, \mathcal{D}) \rightarrow (M(\Omega), \mathcal{D}_{M})$ in a model of knowledge and belief.

Part (\ref{itm:prop_t_i_introspection_1a}) states that player $i$ is certain of her qualitative-type mapping $t_{B_{i}}$ with respect to the possession of beliefs iff her belief operator $B_{i}$ satisfies Positive Introspection. Parts (\ref{itm:prop_t_i_introspection_1a}) and (\ref{itm:prop_t_i_introspection_1b}) jointly state that $B_{i}$ satisfies Positive Introspection and Negative Introspection iff player $i$ is certain of her qualitative-type mapping $t_{B_{i}}$ with respect to $\{ \beta_{E} \mid E \in \mathcal{D} \} \cup \{ \neg \beta_{E} \mid E \in \mathcal{D} \}$. 

I discuss two implications of Proposition \ref{prop:CertaintyIntrospection}. First, it sheds light on the literature of non-partitional knowledge models without Negative Introspection (see footnote \ref{fn:non_partitional}). Part (\ref{itm:prop_t_i_introspection_1}) implies that, without imposing Negative Introspection, player $i$ is not certain of her own qualitative-type mapping with respect to $\mathcal{D}_{M}$ (or $\{ \beta_{E} \mid E \in \mathcal{D} \} \cup \{ \neg \beta_{E} \mid E \in \mathcal{D} \}$). Rather, she takes her own information at face value in the sense that she is only certain of her qualitative-type mapping with respect to her own beliefs $\{ \beta_{E} \mid E \in \mathcal{D} \}$. Proposition \ref{prop:CertaintyIntrospection} formalizes the sense in which \textquotedblleft she takes her own information at face value.'' 

In contrast, Proposition \ref{prop:CertaintyIntrospection} (\ref{itm:prop_t_i_introspection_2a}) shows that, in a partitional possibility correspondence model of knowledge, a player is fully certain of her possibility correspondence when Truth Axiom, (Positive Introspection) and Negative Introspection hold. While the proposition does not necessarily require $B_{i}$ to satisfy the Kripke property, consider a model of knowledge in which $B_{i}$ satisfies Truth Axiom and the Kripke property, i.e., $B_{i}$ is induced by the reflexive possibility correspondence $b_{B_{i}}$. Then, player $i$ is certain of her \textquotedblleft knowledge-generating'' mapping iff $B_{i}$ satisfies (Positive Introspection and) Negative Introspection. 

Proposition \ref{prop:CertaintyIntrospection} (\ref{itm:prop_t_i_introspection_2b}) shows that, for a serial possibility correspondence, a player is fully certain of her possibility correspondence when her beliefs satisfy Positive Introspection and Negative Introspection.

Second, suppose a player has qualitative belief and knowledge. Consider a model $\langle (\Omega, \mathcal{D}), (K_{i})_{i \in I},C \rangle$ where $K_{i}: \mathcal{D} \rightarrow \mathcal{D}$ is player $i$'s (monotone) knowledge operator. Now, for each player $i$, let $B_{i}: \mathcal{D} \rightarrow \mathcal{D}$ be her (monotone) qualitative-belief operator. Let $t_{B_{i}}$ be player $i$'s qualitative-type mapping that represents $B_{i}$, and ask whether player $i$ is certain of her qualitative-type mapping $t_{B_{i}}: (\Omega, \mathcal{D}) \rightarrow (M(\Omega), \mathcal{D}_{M})$. Proposition \ref{prop:CertaintyIntrospection} (\ref{itm:prop_t_i_introspection_2a}) implies that player $i$ is certain of $t_{B_{i}}: (\Omega, \mathcal{D}) \rightarrow (M(\Omega), \mathcal{D}_{M})$ iff $K_{i}$ satisfies \textit{Positive Certainty} (with respect to $B_{i}$): $B_{i}(\cdot) \subseteq K_{i}B_{i}(\cdot)$ and \textit{Negative Certainty} (with respect to $B_{i}$): $(\neg B_{i})(\cdot) \subseteq K_{i}(\neg B_{i})(\cdot)$. Whenever player $i$ believes an event, she knows that she believes it. Whenever player $i$ does not believe an event, she knows that she does not believe it. In fact, these two properties are often assumed in a model of belief and knowledge. Proposition \ref{prop:CertaintyIntrospection} (\ref{itm:prop_t_i_introspection_2a}) justifies the assumptions in terms of the certainty of one's knowledge about her own beliefs.

\section{When are the Players Commonly Certain of a Bleief Model?}\label{sec:ckmodel}

I formalize the sense in which the players are commonly certain of a belief model itself: the players are commonly certain of the profile of their qualitative-type mappings. As discussed, it is sufficient to ask when every player $i$ is certain of each player $j$'s qualitative-type mapping. 

To that end, observe that Proposition \ref{prop:CertaintyIntrospection} applies to the case in which player $i$ is certain of player $j$'s qualitative-type mapping. For example, if player $i$ is certain of player $j$'s qualitative-type mapping $t_{j}: (\Omega, \mathcal{D}) \rightarrow (M(\Omega), \mathcal{D}_{M})$, then $B_{t_{j}}(\cdot) \subseteq B_{i} B_{t_{j}}(\cdot)$ and $(\neg B_{t_{j}})(\cdot) \subseteq B_{i} (\neg B_{t_{j}})(\cdot)$ hold. Proposition \ref{prop:CertaintyIntrospection} implies:

\begin{rmk}\label{rmk:t_j_introspection}
Let $\overrightarrow{\Omega}$ be a belief model, and let $t_{B_{j}}: \Omega \rightarrow M(\Omega)$ be player $j$'s qualitative-type mapping.
\begin{enumerate}
\item \label{itm:prop_t_j_introspection_1} \begin{enumerate}
\item \label{itm:prop_t_j_introspection_1a} Player $i$ is certain of $t_{B_{j}}$ with respect to $\{ \beta_{E} \mid E \in \mathcal{D} \}$ iff $B_{j}(\cdot) \subseteq B_{i}B_{j}(\cdot)$. 
\item \label{itm:prop_t_j_introspection_1b} Player $i$ is certain of $t_{B_{j}}$ with respect to $\{ \neg \beta_{E} \mid E \in \mathcal{D} \}$ iff $(\neg B_{j})(\cdot) \subseteq B_{i} (\neg B_{j})(\cdot)$. 
\item \label{itm:prop_t_j_introspection_1c} If player $i$ is certain of $t_{B_{j}} : (\Omega, \mathcal{D}) \rightarrow (M(\Omega), \mathcal{D}_{M})$, then $B_{j}(\cdot) \subseteq B_{i}B_{j}(\cdot)$ and $(\neg B_{j})(\cdot) \subseteq B_{i} (\neg B_{j})(\cdot)$.
\end{enumerate}
\item \label{itm:prop_t_j_introspection_2} \begin{enumerate}
\item \label{itm:prop_t_j_introspection_2a} Let $B_{i}$ satisfy Truth Axiom. Player $i$ is certain of $t_{B_{j}} : (\Omega, \mathcal{D}) \rightarrow (M(\Omega), \mathcal{D}_{M})$ iff ($B_{j}(\cdot) \subseteq B_{i}B_{j}(\cdot)$ and) $(\neg B_{j})(\cdot) \subseteq B_{i} (\neg B_{j})(\cdot)$. 
\item \label{itm:prop_t_j_introspection_2b} Let $B_{i}$ satisfy Consistency and Countable Conjunction. Player $i$ is certain of $t_{B_{j}} : (\Omega, \mathcal{D}) \rightarrow (M(\Omega), \mathcal{D}_{M})$ iff $B_{j}(\cdot) \subseteq B_{i}B_{j}(\cdot)$ and $(\neg B_{j})(\cdot) \subseteq B_{i} (\neg B_{j})(\cdot)$. 
\end{enumerate}
\end{enumerate}
\end{rmk}

Roughly, Remark \ref{rmk:t_j_introspection} states that player $i$ is certain of player $j$'s qualitative-type mapping $t_{B_{j}}$ if and only if (i) whenever player $j$ believes an event $E$ at $\omega$, player $i$ believes player $j$ believes $E$ at $\omega$; and (ii) whenever player $j$ does not believe an event $E$ at $\omega$, player $i$ believes player $j$ does not believe $E$ at $\omega$. 

Now, I ask when the players are commonly certain of the qualitative-type mappings in a belief model.

\begin{thm}\label{thm:common_certain_qualitative}
Let $\overrightarrow{\Omega}$ be a belief model, and let $t_{B_{i}}: \Omega \rightarrow M(\Omega)$ be player $i$'s qualitative-type mapping. 
\begin{enumerate}
\item \label{itm:common_certain_qualitative_1} Assume Truth Axiom for every $B_{i}$. The players are commonly certain of the profile of qualitative-type mappings $t_{B_{i}} : (\Omega, \mathcal{D}) \rightarrow (M(\Omega), \mathcal{D}_{M})$ iff $B_{i} = B_{j}$ for every $i,j \in I$, (Positive Introspection), and Negative Introspection. In particular, $B_{i}=C$ for each $i \in I$.

\item \label{itm:common_certain_qualitative_2} Assume Consistency and Countable Conjunction for every $B_{i}$. The players are commonly certain of the profile of qualitative-type mappings $t_{B_{i}} : (\Omega, \mathcal{D}) \rightarrow (M(\Omega), \mathcal{D}_{M})$ iff $B_{i}(\cdot) \subseteq C B_{i}(\cdot)$ and $(\neg B_{i})(\cdot) \subseteq C (\neg B_{i})(\cdot)$ for every $i \in I$. In particular, $C = B_{I}$.
\end{enumerate}
\end{thm}

While Part (\ref{itm:common_certain_qualitative_1}) studies a knowledge model, Part (\ref{itm:common_certain_qualitative_2}) does a belief model. I start with discussing implications of Part (\ref{itm:common_certain_qualitative_2}). This part states that the players are commonly certain of their qualitative-type mappings iff (i) for any event $E$ which some player $i$ believes at some state $\omega$, it is commonly believed that player $i$ believes $E$ at $\omega$; and (ii) for any event $E$ which some player $i$ does not believe at some state $\omega$, it is commonly believed that player $i$ does not believe $E$ at $\omega$. 

This part imposes a strong requirement that, under Consistency and Countable Conjunction, the mutual belief and common belief operators coincide if the players are commonly certain of their qualitative-type mappings.\footnote{The converse does not hold, i.e., $C=B_{I}$ does not necessarily imply that the players are certain of the profile of their qualitative-type mappings.} For any event $E$ which everybody believes at some state $\omega$, it is commonly believed that everybody believes $E$ at $\omega$: $B_{I}(\cdot) \subseteq CB_{I}(\cdot)$. Intuitively, in a model of which the players are commonly certain, if everybody believes an event $E$ then it is common belief that everybody believes $E$. Thus, if everybody believes $E$ then everybody believes that everybody believes $E$. Hence, the first-order mutual belief itself implies any higher-order mutual beliefs, and thus the mutual and common beliefs coincide. 

Next, Part (\ref{itm:common_certain_qualitative_1}) provides a contrast between knowledge and belief. In a knowledge model with Truth Axiom, for the players to be commonly certain of the model, it is \textit{necessary} that their knowledge coincides with each other. In contrast, in a belief model without Truth Axiom, there exists a model in which the players' beliefs are different but they are commonly certain of their qualitative-type mappings.

Yet, Theorem \ref{thm:common_certain_qualitative} is an impossibility result in the following sense. In Part (\ref{itm:common_certain_qualitative_1}), every player's knowledge operator coincides. In Part (\ref{itm:common_certain_qualitative_2}), the mutual and common belief operators coincide. In this regard, informally, Theorem \ref{thm:common_certain_qualitative} has some similarity with the impossibility of agreeing-to-disagree \cite{Aumann76}: (under a common prior) if two players have common knowledge of their posteriors then the posteriors coincide. Here, if players are commonly certain of their knowledge operators, then their knowledge operators coincide.  

Finally, as an implication of Theorem \ref{thm:common_certain_qualitative}, suppose that the players are commonly certain of a belief model. If player $i$ is certain of a signal $x$, then is player $j$ certain of the signal $x$, too? While the players' beliefs may not be homogeneous, the proposition below shows that this is the case.

\setcounter{prop}{3}
\begin{prop}\label{prop:common_certainty_signal_implication}
Let $\overrightarrow{\Omega}$ be a belief model such that each $B_{i}$ satisfies Consistency. Let $x: (\Omega, \mathcal{D}) \rightarrow (X, \mathcal{X})$ be a signal such that, for any $F \in \mathcal{X}$, there exists a sub-collection $( F_{\lambda} )_{\lambda \in \Lambda}$ of $\mathcal{X}$ with $F^{c} = \bigcup_{\lambda \in \Lambda} F_{\lambda}$. 
\begin{enumerate}
\item \label{itm:prop_common_certainty_signal_implication_1a} If player $i$ is certain of $x: (\Omega, \mathcal{D}) \rightarrow (X, \mathcal{X})$ and if player $j$ is certain of player $i$'s qualitative-type mapping $t_{B_{i}}: (\Omega, \mathcal{D}) \rightarrow (M(\Omega), \mathcal{D}_{M})$, then player $j$ is certain of $x: (\Omega, \mathcal{D}) \rightarrow (X, \mathcal{X})$.
\item \label{itm:prop_common_certainty_signal_implication_1b} Suppose that the players are commonly certain of the profile of their qualitative-type mappings $t_{B_{i}}: (\Omega, \mathcal{D}) \rightarrow (M(\Omega), \mathcal{D}_{M})$. Then, player $i$ is certain of $x: (\Omega, \mathcal{D}) \rightarrow (X, \mathcal{X})$ iff player $j$ is certain of $x: (\Omega, \mathcal{D}) \rightarrow (X, \mathcal{X})$.
\end{enumerate}
\end{prop}

The meta-common-certainty assumption states that if player $i$ is certain of her own strategy and if player $j$ is certain of player $i$'s type mapping then player $j$ is certain of player $i$'s strategy. In particular, if the players are commonly certain of the profile of their type mappings and if each player is certain of her own strategy, then it follows that the players are commonly certain of the strategy profile. The next section examines the role of such meta-certainty assumptions on game-theoretic solution concepts.

\section{What Role Does the \textquotedblleft Meta-Certainty'' of a Model Play in Game-theoretic Analyses?}\label{sec:solutioncertainty}

This section studies the role that the \textquotedblleft meta-certainty'' assumption plays in game-theoretic analyses of solution concepts. Specifically, it considers the solution concept of iterated elimination of strictly dominated actions (IESDA) in a strategic game. Informally, an epistemic characterization of IESDA states that, in a strategic game, if the (i) \textquotedblleft logical'' players are (ii) \textquotedblleft commonly (meta-)certain of the game'' and if they (iii) commonly believe their rationality, then their resulting actions survive IESDA. Formally, in the context of the framework of this paper, \cite{FukudaCB} shows that if the players commonly believe each player's rationality and if each of them correctly believes their own rationality, then their resulting actions survive IESDA, without assuming any property on individual players' beliefs. This paper connects these two statements as follows: first, suppose that the players are logical in that their beliefs satisfy Consistency and Finite Conjunction in addition to Monotonicity. Second, suppose that each of them is certain of their own qualitative-type mapping and strategy. Third, suppose that the players commonly believe their rationality. Then, their resulting actions survive IESDA.

Here I show that the certainty (of her own strategy and type mapping) allows her to correctly believe her own rationality. In other words, if a player is able to reason about informativeness of her own beliefs, she is able to correctly believe her own rationality.     

\subsection{A Strategic Game, a Model of a Game, and Rationality}

To define the notion of rationality in a game, define a \textit{(strategic) game} as a tuple $\Gamma=\langle (A_{i})_{i \in I}, (\succcurlyeq_{i})_{i \in I} \rangle$: $A_{i}$ is a non-empty at-most-countable set of player $i$'s actions, and $\succcurlyeq_{i}$ is $i$'s (complete and transitive) preference relation on $A := \bigtimes_{i \in I} A_{i}$.\footnote{The assumption on the cardinality of each action set $A_{i}$ is to simplify the analysis. It guarantees that each player is able to reason about any subset of action profiles and that the rationality of each player is an event.} Denote by $\sim_{i}$ and $\succ_{i}$ the indifference and strict relations, respectively. 

A (belief) \textit{model} of the game $\Gamma$ is a tuple $\langle (\Omega, \mathcal{D}), (B_{i})_{i \in I}, C, (\sigma_{i})_{i \in I} \rangle$ (abusing the notation, denote it by $\overrightarrow{\Omega}$) with the following two properties. First, $\langle (\Omega, \mathcal{D}), (B_{i})_{i \in I}, C \rangle$ is a belief model. Second, $\sigma_{i}: \Omega \rightarrow A_{i}$ is a \textit{strategy} of player $i$  satisfying the measurability condition that $\sigma_{i}^{-1}( \{ a_{i} \} ) \in \mathcal{D}$ for all $a_{i} \in A_{i}$. Denote $[\sigma_{i}(\omega)]:= \sigma^{-1}_{i} (\{ \sigma_{i}(\omega) \})$ for each $\omega \in \Omega$.

Denote by $[a'_{i} \succcurlyeq_{i} a_{i}]:=\{ \omega' \in \Omega \mid  (a'_{i}, \sigma_{i}(\omega')) \succcurlyeq_{i} (a_{i}, \sigma_{-i}(\omega')) \} \in \mathcal{D}$ for any $a_{i},a'_{i} \in A_{i}$. In words, $[a'_{i} \succcurlyeq_{i} a_{i}]$ is the event that player $i$ prefers taking action $a'_{i}$ to $a_{i}$ given the opponents' strategies $\sigma_{-i}$. The set $[a'_{i} \succcurlyeq_{i} a_{i}]$ is an event because $[a'_{i} \succcurlyeq_{i} a_{i}] = \sigma_{-i}^{-1}( \{ a_{-i} \in A_{-i} \mid (a'_{i},a_{-i}) \succ_{i} (a_{i},a_{-i}) \} ) \in \mathcal{D}$. Define $[a'_{i} \succ_{i} a_{i}]$ and $[a'_{i} \sim_{i} a_{i}]$ analogously.

Denote by $\mathrm{RAT}_{i}$ the event that player $i$ is \textit{rational} (see, e.g., \cite{BonannoLOFT08, BonannoHandbook15, ChenGEB07}): 
\begin{align*}
\mathrm{RAT}_{i} := & \{ \omega \in \Omega \mid \omega \in B_{i}( [a'_{i} \succ_{i} \sigma_{i}(\omega)]  ) \text{ for no } a'_{i} \in A_{i} \}.
\end{align*}
It can be seen that $\mathrm{RAT}_{i}$ is indeed an event. Let $\mathrm{RAT}_{I}:=\bigcap_{i \in I} \mathrm{RAT}_{i}$. Player $i$ is \textit{rational} at $\omega \in \Omega$ if there is no action $a'_{i} \in A_{i}$ such that she believes that playing $a'_{i}$ is strictly better than playing $\sigma_{i}(\omega)$ given the opponents' strategies $\sigma_{-i}$. In other words, player $i$ is rational at $\omega$ if, for any action $a'_{i}$, she always considers it possible that playing $\sigma_{i}(\omega)$ is at least as good as playing $a'_{i}$ given the opponents' strategies $\sigma_{-i}$: $\omega \in (\neg B_{i})( \neg [\sigma_{i}(\omega) \succcurlyeq_{i} a'_{i}] )$ for any $a'_{i} \in A_{i}$. 

Now, the epistemic characterization of IESDA is stated as follows. Suppose that each player $i$ correctly believes her own rationality: $B_{i}(\mathrm{RAT}_{i}) \subseteq \mathrm{RAT}_{i}$. If every player's rationality is common belief at $\omega$, i.e., $\omega \in \bigcap_{i \in I} C(\mathrm{RAT}_{i})$, then the resulting actions $(\sigma_{i}(\omega))_{i \in I} \in A$ survive any process of IESDA.\footnote{Since each player's belief operator $B_{i}$ satisfies Monotonicity, $B_{i}(\mathrm{RAT}_{I}) \subseteq \bigcap_{i \in I} B_{i}(\mathrm{RAT}_{i})$ and $C(\mathrm{RAT}_{I}) \subseteq \bigcap_{i \in I} C(\mathrm{RAT}_{i})$. Thus, if every player $i$ correctly believes the rationality of the players, then each player correctly believes her own rationality. Likewise, if it is common belief that the players are rational, then, for every $i \in I$, it is common belief that player $i$ is rational. Hence, I examine the weaker condition that each player $i$ correctly believes her own rationality.} 

Finally, player $i$ is \textit{certain of her own strategy} $\sigma_{i}$ if she is certain of $\sigma_{i}: (\Omega, \mathcal{D}) \rightarrow (A_{i}, \{ \{ a_{i} \} \mid a_{i} \in A_{i} \})$, equivalently, $[\sigma_{i}(\cdot)] \subseteq B_{i}([\sigma_{i}(\cdot)])$. Note that, under Consistency in addition to Monotonicity, if player $i$ is certain of her own strategy then $B_{i}([\sigma_{i}(\cdot)]) = [\sigma_{i}(\cdot)]$, $[\sigma_{i}(\cdot)]^{c} = B_{i}([\sigma_{i}(\cdot)]^{c})$, and $B_{i}(\Omega) = \Omega$.\footnote{Thus, under Consistency and Monotonicity of $B_{i}$, the certainty of own strategy implies that if player $i$ is rational at $\omega$, then she never takes a strictly dominated action at $\omega$ (if she takes a strictly dominated action, then her belief violates Necessitation).} 

\subsection{The Role of Meta-certainty in Correctly Believing One's Own Rationality}

I ask under what conditions player $i$ correctly believes her own rationality: $B_{i}(\mathrm{RAT}_{i}) \subseteq \mathrm{RAT}_{i}$. For qualitative belief, the standard assumptions on qualitative belief (i.e., Consistency, Positive Introspection, Negative Introspection, and the Kripke property) guarantee that $B_{i}(\mathrm{RAT}_{i}) = \mathrm{RAT}_{i}$ (e.g., \cite{BonannoLOFT08, BonannoHandbook15}).\footnote{It can be seen that Consistency, Positive Introspection, and the Kripke property in addition to the certainty of $i$'s own strategy yield $B_{i}(\mathrm{RAT}_{i}) \subseteq \mathrm{RAT}_{i}$. Likewise, Negative Introspection and the Kripke property in addition to the certainty of $i$'s own strategy yield $\mathrm{RAT}_{i} \subseteq B_{i}(\mathrm{RAT}_{i})$.} Here, I provide a compatibility condition on belief with informativeness, under which a player correctly believes her own rationality. The compatibility condition does not hinge on a particular form of belief, i.e., whether it is qualitative or probabilistic. 

To that end, a state $\omega$ is \textit{at least as informative as} another state $\omega'$ to $i$ (precisely, according to $t_{B_{i}}$) iff $t_{B_{i}}(\omega')(\cdot) \leq t_{B_{i}}(\omega)(\cdot)$. Fix $\omega \in \Omega$, and let $(\uparrow t_{B_{i}}(\omega) ) := \{ \omega' \in \Omega \mid t_{B_{i}}(\omega)(\cdot) \leq t_{B_{i}}(\omega')(\cdot) \}$ be the set of states that are at least as informative to $i$ as $\omega$. Under the Kripke property, $\omega' \in (\uparrow t_{B_{i}}(\omega) )$ iff $b_{B_{i}}(\omega') \subseteq b_{B_{i}}(\omega)$. The full paper extensively studies the notion of informativeness. Now:

\begin{defi}\label{df:compatible_informativeness}
Player $i$'s belief (operator $B_{i}$) is \textit{compatible with informativeness} if $(\uparrow t_{B_{i}}(\omega)) \cap E \neq \emptyset$ for any $E \in \mathcal{D}$ with $\omega \in B_{i}(E)$. 
\end{defi}

In words, player $i$'s beliefs are compatible with informativeness if, for any event $E$ which player $i$ believes at some $\omega$, there exists a state $\omega'$ in $E$ which is at least as informative as $\omega$. In the context of qualitative beliefs, if player $i$'s belief operator $B_{i}$ satisfies the Kripke property, Consistency, and Positive Introspection, then $B_{i}$ is compatible with informativeness. The compatibility with informativeness does not necessarily imply the Kripke property (and vice versa). If player $i$'s belief operator $B_{i}$ is compatible with informativeness, then it satisfies $B_{i}(\emptyset)=\emptyset$. Thus, under Finite Conjunction, if $B_{i}$ is compatible with informativeness, then it satisfies Consistency. 

The following proposition states that the compatibility of beliefs with informativeness is implied by the certainty of a type mapping. 

\begin{prop}\label{prop:compatible_informative}
Let $\overrightarrow{\Omega}$ be a belief model. Assume: (i) $(\uparrow t_{B_{i}}(\cdot)) \in \mathcal{D}$; (ii) $B_{i}$ satisfies Consistency and Finite Conjunction; and that (iii) player $i$ is certain of $t_{B_{i}}: \Omega \rightarrow M(\Omega)$ with respect to $\{ \{ \mu \in M(\Omega) \mid \mu(\cdot) \geq t_{B_{i}}(\omega)(\cdot) \} \mid \omega \in \Omega \}$. Then, $B_{i}$ is compatible with informativeness.
\end{prop}

The proposition states that, under the regularity condition (i), if player $i$ is logical (in that her belief operator satisfies Consistency and Finite Conjunction) and if she is certain of her qualitative-type mapping, then her beliefs are compatible with informativeness. Theorem \ref{thm:informativeness_rationality} below establishes that if player $i$'s beliefs are compatible with informativeness then she correctly believes her rationality, which is a part of the preconditions of the epistemic characterization of IESDA.

Now, the main result of this section is as follows: a player correctly believes her own rationality if: (i) she is certain of her own strategy; (ii) her belief is compatible with the informativeness; and if (iii) her belief is (finitely) conjunctive so that she can simultaneously reason about her own strategy and her own rationality.

\setcounter{thm}{1}
\begin{thm}\label{thm:informativeness_rationality}
Suppose that player $i$ is certain of her own strategy (i.e., $[\sigma_{i}(\cdot)] \subseteq B_{i}([\sigma_{i}(\cdot)])$). Also, let $B_{i}$ be compatible with informativeness and satisfy Finite Conjunction. Then, player $i$ correctly believes her own rationality: $B_{i}(\mathrm{RAT}_{i}) \subseteq \mathrm{RAT}_{i}$.
\end{thm}

Proposition \ref{prop:compatible_informative} and Theorem \ref{thm:informativeness_rationality} imply that player $i$ correctly believes her own rationality if she is logical in that her belief operator satisfies Consistency and Finite Conjunction and if she is certain of her own type mapping and strategy. Theorem \ref{thm:informativeness_rationality} states that, for the role of the meta-certainty assumption of a belief model on IESDA, it is not necessary that each player is certain of the profile of type mappings but it is sufficient that each player is certain of her own type mapping. One can incorporate the assumptions that each player is certain of her own qualitative type-mapping and strategy into the condition that she is certain of the part of the model of a game $\langle (\Omega, \mathcal{D}), (t_{B_{i}}, \sigma_{i}) \rangle$ that dictates her beliefs and strategy. 


\bibliographystyle{eptcs}
\bibliography{TARK2021bib}

\begin{thebibliography}{10}
\providecommand{\bibitemdeclare}[2]{}
\providecommand{\surnamestart}{}
\providecommand{\surnameend}{}
\providecommand{\urlprefix}{Available at }
\providecommand{\url}[1]{\texttt{#1}}
\providecommand{\href}[2]{\texttt{#2}}
\providecommand{\urlalt}[2]{\href{#1}{#2}}
\providecommand{\doi}[1]{doi:\urlalt{http://dx.doi.org/#1}{#1}}
\providecommand{\bibinfo}[2]{#2}

\bibitemdeclare{article}{Aumann76}
\bibitem{Aumann76}
\bibinfo{author}{Robert~J. \surnamestart Aumann\surnameend}
  (\bibinfo{year}{1976}): \emph{\bibinfo{title}{Agreeing to Disagree}}.
\newblock {\sl \bibinfo{journal}{The Annals of Statistics}}
  \bibinfo{volume}{4}(\bibinfo{number}{6}), pp. \bibinfo{pages}{1236--1239},
  \doi{10.1214/aos/1176343654}.

\bibitemdeclare{article}{Aumann87}
\bibitem{Aumann87}
\bibinfo{author}{Robert~J. \surnamestart Aumann\surnameend}
  (\bibinfo{year}{1987}): \emph{\bibinfo{title}{Correlated Equilibrium as an
  Expression of Bayesian Rationality}}.
\newblock {\sl \bibinfo{journal}{Econometrica}}
  \bibinfo{volume}{55}(\bibinfo{number}{1}), pp. \bibinfo{pages}{1--18},
  \doi{10.2307/1911154}.

\bibitemdeclare{article}{Aumann99}
\bibitem{Aumann99}
\bibinfo{author}{Robert~J. \surnamestart Aumann\surnameend}
  (\bibinfo{year}{1999}): \emph{\bibinfo{title}{Interactive Epistemology I,
  II}}.
\newblock {\sl \bibinfo{journal}{International Journal of Game Theory}}
  \bibinfo{volume}{28}(\bibinfo{number}{3}), pp. \bibinfo{pages}{263--300,
  301--314}.
\newblock
  \bibinfo{note}{\href{http://doi.org/10.1007/s001820050111}{doi:10.1007/s001820050111},
  \href{http://doi.org/10.1007/s001820050112}{doi:10.1007/s001820050112}}.

\bibitemdeclare{article}{Bacharach85}
\bibitem{Bacharach85}
\bibinfo{author}{Michael \surnamestart Bacharach\surnameend}
  (\bibinfo{year}{1985}): \emph{\bibinfo{title}{Some Extensions of a Claim of
  Aumann in an Axiomatic Model of Knowledge}}.
\newblock {\sl \bibinfo{journal}{Journal of Economic Theory}}
  \bibinfo{volume}{37}(\bibinfo{number}{1}), pp. \bibinfo{pages}{167--190},
  \doi{10.1016/0022-0531(85)90035-3}.

\bibitemdeclare{incollection}{Bacharach90}
\bibitem{Bacharach90}
\bibinfo{author}{Michael \surnamestart Bacharach\surnameend}
  (\bibinfo{year}{1990}): \emph{\bibinfo{title}{When Do We Have Information
  Partitions?}}
\newblock In \bibinfo{editor}{M.~\surnamestart Bacharach\surnameend} \&
  \bibinfo{editor}{M.~\surnamestart Dempster\surnameend}, editors: {\sl
  \bibinfo{booktitle}{Mathematical Models in Economics}},
  \bibinfo{publisher}{Oxford University Press}, pp. \bibinfo{pages}{1--23}.

\bibitemdeclare{incollection}{BinmoreBrandenburger}
\bibitem{BinmoreBrandenburger}
\bibinfo{author}{K.~\surnamestart Binmore\surnameend} \&
  \bibinfo{author}{A.~\surnamestart Brandenburger\surnameend}
  (\bibinfo{year}{1990}): \emph{\bibinfo{title}{Common Knowledge and Game
  Theory}}.
\newblock In \bibinfo{editor}{K.~\surnamestart Binmore\surnameend}, editor:
  {\sl \bibinfo{booktitle}{Essays on the Foundations of Game Theory}},
  \bibinfo{publisher}{Basil Blackwell}, pp. \bibinfo{pages}{105--150}.

\bibitemdeclare{incollection}{BonannoLOFT08}
\bibitem{BonannoLOFT08}
\bibinfo{author}{Giacomo \surnamestart Bonanno\surnameend}
  (\bibinfo{year}{2008}): \emph{\bibinfo{title}{A Syntactic Approach to
  Rationality in Games with Ordinal Payoffs}}.
\newblock In \bibinfo{editor}{Giacomo \surnamestart Bonanno\surnameend},
  \bibinfo{editor}{Wiebe \surnamestart van~der Hoek\surnameend} \&
  \bibinfo{editor}{Michael \surnamestart Wooldridge\surnameend}, editors: {\sl
  \bibinfo{booktitle}{Logic and the Foundations of Game and Decision Theory
  (LOFT 7)}}, \bibinfo{publisher}{Amsterdam University Press}, pp.
  \bibinfo{pages}{59--86}, \doi{10.5117/9789089640260}.

\bibitemdeclare{incollection}{BonannoHandbook15}
\bibitem{BonannoHandbook15}
\bibinfo{author}{Giacomo \surnamestart Bonanno\surnameend}
  (\bibinfo{year}{2015}): \emph{\bibinfo{title}{Epistemic Foundations of Game
  Theory}}.
\newblock In \bibinfo{editor}{Hans \surnamestart van Ditmarsch\surnameend},
  \bibinfo{editor}{Joseph~Y. \surnamestart Halpern\surnameend},
  \bibinfo{editor}{Wiebe \surnamestart van~der Hoek\surnameend} \&
  \bibinfo{editor}{Barteld~Pieter \surnamestart Kooi\surnameend}, editors: {\sl
  \bibinfo{booktitle}{Handbook of Epistemic Logic}},
  \bibinfo{publisher}{College Publications}, pp. \bibinfo{pages}{443--487}.

\bibitemdeclare{article}{BrandenburgerDekelEcta87}
\bibitem{BrandenburgerDekelEcta87}
\bibinfo{author}{Adam \surnamestart Brandenburger\surnameend} \&
  \bibinfo{author}{Eddie \surnamestart Dekel\surnameend}
  (\bibinfo{year}{1987}): \emph{\bibinfo{title}{Rationalizability and
  Correlated Equilibria}}.
\newblock {\sl \bibinfo{journal}{Econometrica}}
  \bibinfo{volume}{55}(\bibinfo{number}{6}), pp. \bibinfo{pages}{1391--1402},
  \doi{10.2307/1913562}.

\bibitemdeclare{incollection}{BDCK}
\bibitem{BDCK}
\bibinfo{author}{Adam \surnamestart Brandenburger\surnameend} \&
  \bibinfo{author}{Eddie \surnamestart Dekel\surnameend}
  (\bibinfo{year}{1989}): \emph{\bibinfo{title}{The Role of Common Knowledge
  Assumptions in Game Theory}}.
\newblock In \bibinfo{editor}{Frank \surnamestart Hahn\surnameend}, editor:
  {\sl \bibinfo{booktitle}{The Economics of Missing Markets, Information, and
  Games}}, \bibinfo{publisher}{Oxford University Press}, pp.
  \bibinfo{pages}{46--61}.

\bibitemdeclare{article}{BDHierarchies}
\bibitem{BDHierarchies}
\bibinfo{author}{Adam \surnamestart Brandenburger\surnameend} \&
  \bibinfo{author}{Eddie \surnamestart Dekel\surnameend}
  (\bibinfo{year}{1993}): \emph{\bibinfo{title}{Hierarchies of Beliefs and
  Common Knowledge}}.
\newblock {\sl \bibinfo{journal}{Journal of Economic Theory}}
  \bibinfo{volume}{59}(\bibinfo{number}{1}), pp. \bibinfo{pages}{189--198},
  \doi{10.1006/jeth.1993.1012}.

\bibitemdeclare{article}{BDG}
\bibitem{BDG}
\bibinfo{author}{Adam \surnamestart Brandenburger\surnameend},
  \bibinfo{author}{Eddie \surnamestart Dekel\surnameend} \&
  \bibinfo{author}{John \surnamestart Geanakoplos\surnameend}
  (\bibinfo{year}{1992}): \emph{\bibinfo{title}{Correlated Equilibrium with
  Generalized Information Structures}}.
\newblock {\sl \bibinfo{journal}{Games and Economic Behavior}}
  \bibinfo{volume}{4}(\bibinfo{number}{2}), pp. \bibinfo{pages}{182--201},
  \doi{10.1016/0899-8256(92)90014-J}.

\bibitemdeclare{article}{BKImpossibility}
\bibitem{BKImpossibility}
\bibinfo{author}{Adam \surnamestart Brandenburger\surnameend} \&
  \bibinfo{author}{H.~Jerome \surnamestart Keisler\surnameend}
  (\bibinfo{year}{2006}): \emph{\bibinfo{title}{An Impossibility Theorem on
  Beliefs in Games}}.
\newblock {\sl \bibinfo{journal}{Studia Logica}}
  \bibinfo{volume}{84}(\bibinfo{number}{2}), pp. \bibinfo{pages}{211--240},
  \doi{10.1007/s11225-006-9011-z}.

\bibitemdeclare{article}{ChenGEB07}
\bibitem{ChenGEB07}
\bibinfo{author}{Yi-Chun \surnamestart Chen\surnameend},
  \bibinfo{author}{Ngo~Van \surnamestart Long\surnameend} \&
  \bibinfo{author}{Xiao \surnamestart Luo\surnameend} (\bibinfo{year}{2007}):
  \emph{\bibinfo{title}{Iterated Strict Dominance in General Games}}.
\newblock {\sl \bibinfo{journal}{Games and Economic Behavior}}
  \bibinfo{volume}{61}(\bibinfo{number}{2}), pp. \bibinfo{pages}{299--315},
  \doi{10.1016/j.geb.2007.02.002}.

\bibitemdeclare{incollection}{DekelGul}
\bibitem{DekelGul}
\bibinfo{author}{Eddie \surnamestart Dekel\surnameend} \&
  \bibinfo{author}{Faruk \surnamestart Gul\surnameend} (\bibinfo{year}{1997}):
  \emph{\bibinfo{title}{Rationality and Knowledge in Game Theory}}.
\newblock In \bibinfo{editor}{David~M. \surnamestart Kreps\surnameend} \&
  \bibinfo{editor}{Kenneth~F. \surnamestart Wallis\surnameend}, editors: {\sl
  \bibinfo{booktitle}{Advances in Economics and Econometrics: Theory and
  Applications, Seventh World Congress}}, \bibinfo{volume}{1},
  \bibinfo{publisher}{Cambridge University Press}, pp.
  \bibinfo{pages}{87--172}, \doi{10.1017/CCOL521580110.005}.

\bibitemdeclare{article}{FGHV}
\bibitem{FGHV}
\bibinfo{author}{Ronald \surnamestart Fagin\surnameend}, \bibinfo{author}{John
  \surnamestart Geanakoplos\surnameend}, \bibinfo{author}{Joseph~Y.
  \surnamestart Halpern\surnameend} \& \bibinfo{author}{Moshe~Y. \surnamestart
  Vardi\surnameend} (\bibinfo{year}{1999}): \emph{\bibinfo{title}{The
  Hierarchical Approach to Modeling Knowledge and Common Knowledge}}.
\newblock {\sl \bibinfo{journal}{International Journal of Game Theory}}
  \bibinfo{volume}{28}(\bibinfo{number}{3}), pp. \bibinfo{pages}{331--365},
  \doi{10.1007/s001820050114}.

\bibitemdeclare{incollection}{FukudaKnowledgeChap1}
\bibitem{FukudaKnowledgeChap1}
\bibinfo{author}{Satoshi \surnamestart Fukuda\surnameend}
  (\bibinfo{year}{2017}): \emph{\bibinfo{title}{The Existence of Universal
  Knowledge Spaces}}.
\newblock In: {\sl \bibinfo{booktitle}{Essays in the Economics of Information
  and Epistemology}}, \bibinfo{publisher}{Ph.D. Dissertation, the University of
  California at Berkeley}, pp. \bibinfo{pages}{1--113}.

\bibitemdeclare{article}{FukudaCB}
\bibitem{FukudaCB}
\bibinfo{author}{Satoshi \surnamestart Fukuda\surnameend}
  (\bibinfo{year}{2020}): \emph{\bibinfo{title}{Formalizing Common Belief with
  No Underlying Assumption on Individual Beliefs}}.
\newblock {\sl \bibinfo{journal}{Games and Economic Behavior}}
  \bibinfo{volume}{121}, pp. \bibinfo{pages}{169--189},
  \doi{10.1016/j.geb.2020.02.007}.

\bibitemdeclare{unpublished}{FukudaUQB}
\bibitem{FukudaUQB}
\bibinfo{author}{Satoshi \surnamestart Fukuda\surnameend}
  (\bibinfo{year}{2021}): \emph{\bibinfo{title}{The Existence of Universal
  Qualitative Belief Spaces}}.

\bibitemdeclare{techreport}{Geanakoplos}
\bibitem{Geanakoplos}
\bibinfo{author}{John \surnamestart Geanakoplos\surnameend}
  (\bibinfo{year}{1989}): \emph{\bibinfo{title}{Game Theory without Partitions,
  and Applications to Speculation and Consensus}}.
\newblock \bibinfo{type}{Cowles Foundation Discussion Paper No. 914, Yale
  University}.

\bibitemdeclare{conference}{GilboaMeta}
\bibitem{GilboaMeta}
\bibinfo{author}{Itzhak \surnamestart Gilboa\surnameend}
  (\bibinfo{year}{1988}): \emph{\bibinfo{title}{Information and
  Meta-Information}}.
\newblock In: {\sl \bibinfo{booktitle}{Proceedings of the 2nd Conference on
  Theoretical Aspects of Reasoning about Knowledge}},
  \bibinfo{publisher}{Morgan Kaufmann Publishers Inc.}, pp.
  \bibinfo{pages}{227--243}.

\bibitemdeclare{article}{Harsanyi}
\bibitem{Harsanyi}
\bibinfo{author}{John~C. \surnamestart Harsanyi\surnameend}
  (\bibinfo{year}{1967-1968}): \emph{\bibinfo{title}{Games with Incomplete
  Information Played by \textquotedblleft Bayesian'' Players, I-III}}.
\newblock {\sl \bibinfo{journal}{Management Science}} \bibinfo{volume}{14}, pp.
  \bibinfo{pages}{159--182, 320--334, 486--502}.
\newblock
  \bibinfo{note}{\href{http://doi.org/10.1287/mnsc.14.3.159}{doi:10.1287/mnsc.14.3.159},
  \href{http://doi.org/10.1287/mnsc.14.5.320}{doi:10.1287/mnsc.14.5.320},
  \href{http://doi.org/10.1287/mnsc.14.7.486}{doi:10.1287/mnsc.14.7.486}}.

\bibitemdeclare{article}{MondererSamet}
\bibitem{MondererSamet}
\bibinfo{author}{Dov \surnamestart Monderer\surnameend} \& \bibinfo{author}{Dov
  \surnamestart Samet\surnameend} (\bibinfo{year}{1989}):
  \emph{\bibinfo{title}{Approximating Common Knowledge with Common Beliefs}}.
\newblock {\sl \bibinfo{journal}{Games and Economic Behavior}}
  \bibinfo{volume}{1}(\bibinfo{number}{2}), pp. \bibinfo{pages}{170--190},
  \doi{10.1016/0899-8256(89)90017-1}.

\bibitemdeclare{article}{MorrisJET}
\bibitem{MorrisJET}
\bibinfo{author}{Stephen \surnamestart Morris\surnameend}
  (\bibinfo{year}{1996}): \emph{\bibinfo{title}{The Logic of Belief and Belief
  Change: A Decision Theoretic Approach}}.
\newblock {\sl \bibinfo{journal}{Journal of Economic Theory}}
  \bibinfo{volume}{69}(\bibinfo{number}{1}), pp. \bibinfo{pages}{1--23},
  \doi{10.1006/jeth.1996.0035}.

\bibitemdeclare{unpublished}{Pires}
\bibitem{Pires}
\bibinfo{author}{Cesaltina~Pacheco \surnamestart Pires\surnameend}
  (\bibinfo{year}{1994}): \emph{\bibinfo{title}{Do I know $\Omega$? An
  Axiomatic Model of Awareness and Knowledge}}.

\bibitemdeclare{article}{RoyPacuit}
\bibitem{RoyPacuit}
\bibinfo{author}{Olivier \surnamestart Roy\surnameend} \& \bibinfo{author}{Eric
  \surnamestart Pacuit\surnameend} (\bibinfo{year}{2013}):
  \emph{\bibinfo{title}{Substantive assumptions in interaction: a logical
  perspective}}.
\newblock {\sl \bibinfo{journal}{Synthese}}
  \bibinfo{volume}{190}(\bibinfo{number}{5}), pp. \bibinfo{pages}{891--908},
  \doi{10.1007/s11229-012-0191-y}.

\bibitemdeclare{article}{SametIgnorance}
\bibitem{SametIgnorance}
\bibinfo{author}{Dov \surnamestart Samet\surnameend} (\bibinfo{year}{1990}):
  \emph{\bibinfo{title}{Ignoring Ignorance and Agreeing to Disagree}}.
\newblock {\sl \bibinfo{journal}{Journal of Economic Theory}}
  \bibinfo{volume}{52}(\bibinfo{number}{1}), pp. \bibinfo{pages}{190--207},
  \doi{10.1016/0022-0531(90)90074-T}.

\bibitemdeclare{article}{Shin}
\bibitem{Shin}
\bibinfo{author}{Hyun~Song \surnamestart Shin\surnameend}
  (\bibinfo{year}{1993}): \emph{\bibinfo{title}{Logical Structure of Common
  Knowledge}}.
\newblock {\sl \bibinfo{journal}{Journal of Economic Theory}}
  \bibinfo{volume}{60}(\bibinfo{number}{1}), pp. \bibinfo{pages}{1--13},
  \doi{10.1006/jeth.1993.1032}.

\bibitemdeclare{article}{StalnakerTD94}
\bibitem{StalnakerTD94}
\bibinfo{author}{Robert \surnamestart Stalnaker\surnameend}
  (\bibinfo{year}{1994}): \emph{\bibinfo{title}{On the Evaluation of Solution
  Concepts}}.
\newblock {\sl \bibinfo{journal}{Theory and Decision}}
  \bibinfo{volume}{37}(\bibinfo{number}{1}), pp. \bibinfo{pages}{49--73},
  \doi{10.1007/BF01079205}.

\bibitemdeclare{article}{TanWerlangJET}
\bibitem{TanWerlangJET}
\bibinfo{author}{Tommy Chin-Chiu \surnamestart Tan\surnameend} \&
  \bibinfo{author}{S\'{e}rgio Ribeiro da~Costa \surnamestart
  Werlang\surnameend} (\bibinfo{year}{1988}): \emph{\bibinfo{title}{The
  Bayesian Foundations of Solution Concepts of Games}}.
\newblock {\sl \bibinfo{journal}{Journal of Economic Theory}}
  \bibinfo{volume}{45}(\bibinfo{number}{2}), pp. \bibinfo{pages}{370--391},
  \doi{10.1016/0022-0531(88)90276-1}.

\bibitemdeclare{article}{TanWerlangRBE}
\bibitem{TanWerlangRBE}
\bibinfo{author}{Tommy Chin-Chiu \surnamestart Tan\surnameend} \&
  \bibinfo{author}{S\'{e}rgio Ribeiro da~Costa \surnamestart
  Werlang\surnameend} (\bibinfo{year}{1992}): \emph{\bibinfo{title}{On Aumann's
  Notion of Common Knowledge: an Alternative Approach}}.
\newblock {\sl \bibinfo{journal}{Revista Brasileira de Economia - RBE}}
  \bibinfo{volume}{46}(\bibinfo{number}{2}), pp. \bibinfo{pages}{151--166}.

\bibitemdeclare{article}{VassilakisZamir}
\bibitem{VassilakisZamir}
\bibinfo{author}{Spyros \surnamestart Vassilakis\surnameend} \&
  \bibinfo{author}{Shmuel \surnamestart Zamir\surnameend}
  (\bibinfo{year}{1993}): \emph{\bibinfo{title}{Common Belief and Common
  Knowledge}}.
\newblock {\sl \bibinfo{journal}{Journal of Mathematical Economics}}
  \bibinfo{volume}{22}(\bibinfo{number}{5}), pp. \bibinfo{pages}{495--505},
  \doi{10.1016/0304-4068(93)90039-N}.

\bibitemdeclare{incollection}{WerlangCKDict}
\bibitem{WerlangCKDict}
\bibinfo{author}{S\'{e}rgio Ribeiro da~Costa \surnamestart Werlang\surnameend}
  (\bibinfo{year}{1987}): \emph{\bibinfo{title}{Common Knowledge}}.
\newblock In \bibinfo{editor}{John \surnamestart Eatwell\surnameend},
  \bibinfo{editor}{Murray \surnamestart Milgate\surnameend} \&
  \bibinfo{editor}{Peter \surnamestart Newman~K.\surnameend}, editors: {\sl
  \bibinfo{booktitle}{The New Palgrave: A Dictionary of Economics}},
  \bibinfo{edition}{2nd} edition, \bibinfo{publisher}{Macmillan}, pp.
  \bibinfo{pages}{74--85}, \doi{10.1007/978-1-349-20181-5_5}.

\bibitemdeclare{incollection}{Wilson87}
\bibitem{Wilson87}
\bibinfo{author}{Robert \surnamestart Wilson\surnameend}
  (\bibinfo{year}{1987}): \emph{\bibinfo{title}{Game-Theoretic Analyses of
  Trading Processes}}.
\newblock In \bibinfo{editor}{Truman~Fassett \surnamestart Bewley\surnameend},
  editor: {\sl \bibinfo{booktitle}{Advances in Economic Theory: Fifth World
  Congress}}, \bibinfo{publisher}{Cambridge University Press}, pp.
  \bibinfo{pages}{33--70}, \doi{10.1017/CCOL0521340446.002}.

\end{thebibliography}

\end{document}